\documentclass{article}
\usepackage{emulateapj}
\usepackage{amsmath}
\usepackage{amssym}
\usepackage{psfig}
\usepackage{times}

\def\Ep{E_{\rm p}}
\def\NE{N_{\rm E}}
\def\E0{E_{\rm 0}}
\def\P0{\Phi_{\rm 0}}
\def\Ep0{E_{\rm p0}}

\def\fEEpk{f_{\rm E} (E/ E _{\rm pk}, E _{\rm pk})}
\def\fxEpk{f_{\rm x} ( x, E _{\rm pk})}
\def\ai{\alpha_{\rm i}}
\def\bi{\beta_{\rm i}}
\def\ati{\alpha_{\rm ti}}
\def\bti{\beta_{\rm ti}}

\newcommand{\bez}{\begin{eqnarray*}}
\newcommand{\eez}{\end{eqnarray*}}
\newcommand{\be}{\begin{equation}}
\newcommand{\ee}{\end{equation}}
\newcommand{\beq}{\begin{eqnarray}}
\newcommand{\eeq}{\end{eqnarray}}
\newcommand{\bc}{\begin{center}}
\newcommand{\ec}{\end{center}}

\slugcomment{to appear in the ApJ}

\lefthead{Ryde \& Svensson}
\righthead{On the Shape of Pulse Spectra in GRBs }

\begin{document}
 
\title{On the Shape of Pulse Spectra in Gamma-Ray Bursts}
\author{Felix Ryde and Roland Svensson}
\affil{Stockholm Observatory, S-133 36 Saltsj\"obaden, Sweden}

\begin{abstract}
The discovery (Liang \& Kargatis 1996), that the 
peak energy of time-resolved spectra of gamma-ray burst (GRB)
pulses
decays exponentially with fluence, is analytically shown to imply
that the time-integrated photon number spectrum of a pulse should have a 
unique shape, given by an underlying $E ^{-1}$ behavior.
We also show that the asymptotic low energy normalization
of the time-integrated spectrum is equal to the exponential decay constant.
We study analytically how this general behavior is modified
in more realistic situations and show that diversity is then introduced
in the properties of time-integrated GRB pulse spectra.
We argue that further diversity will occur in 
time-integrated multi-pulse (complex) GRB spectra.
The total energy received per cm$^2$ is approximately the decay constant
times the maximum peak energy of the pulse.
Our analytical results connect the properties of the
time-integrated pulse spectrum with those of the time-resolved
spectra, and can thus be used when studying observed GRB 
pulse spectra.  
We illustrate with the bright burst GRB 910807 and comment on GRB 910525 and
GRB 921207. 
\end{abstract}


\section{Introduction}
The time-integrated and time-resolved spectra of gamma-ray bursts 
(GRB) differ.
The general appearance of time-integrated GRB spectra is that of
a smoothly broken power law. 
Time-integrated photon number spectra from missions prior to the  
{\it Compton Gamma-Ray Observatory (CGRO)} with its Burst and 
Transient Source Experiment (BATSE) 
were often fitted with an ``optically thin thermal bremsstrahlung''
(OTTB) spectrum, ${\cal {N}} _{\rm E} (E)\propto E^{-1} e^{-E/E_{0}}$, 
combined with a power law at larger energies. 
Band et al. (1993) showed that nearly all time-integrated BATSE spectra were
well fitted by the ``Band'' function consisting of two power laws, 
${\cal {N}}  _{\rm E} (E) \propto E ^\alpha$ at low energies and 
${\cal {N}} _{\rm E} (E) \propto E ^\beta$ at 
high energies,
smoothly joined by an exponential roll-over characterized by a 
constant  $E_0$.  
The Band function index, $\alpha$, shows
a broad peak between -1.25 and -0.25 (Band et~al. 1993), 
instead of the earlier claimed universal value of -1,
$\beta$ clusters fairly narrowly around -2.1 (Preece et~al. 1997), while 
$E_0$ has values between 100 keV and 1 MeV, but mainly
clustering around 100-200 keV (Band et~al. 1993).

Time-resolved spectra, even within single pulses,
show strong time evolution, and are, in general, harder than 
time-integrated spectra ({\it e.g.}, Pendleton et~al. 1994a,
Ford et~al. 1995, 
Liang \& Kargatis 1996, hereafter LK96,
Crider et~al. 1997a, 1997b). Most importantly,
LK96 found  
that the peak energy, $E_{\rm pk}$, 
of the time-resolved $E^2 {\cal {N}} _{\rm E} (E)$ spectra of a single 
pulse in a number of bright and smooth GRBs, decays
exponentially  as a function of photon fluence, {\it i.e.},
\begin{equation}
E_{\rm pk} (t) = E_{\rm pk,max} e ^{-\Phi (t) / \Phi_{\rm 0}}.
\end{equation} 
\noindent
where $E_{\rm pk,max}$ is the maximum value of $E_{\rm pk}$ 
within a pulse,
$\Phi (t)$ is the photon fluence integrated from the time of 
$E_{\rm pk,max}$, and $\Phi _{\rm 0}$
is the exponential decay constant. 
Furthermore, Crider et al. (1997a) found the instantaneous 
Band function index, $\ai$,  in general to be hard, 
$> -1$, reaching values as large as +1.6.

The aim of this paper is to show the consequences
of the time evolution of $E_{\rm pk} (t)$, for 
the time-integrated GRB spectrum of a single pulse, 
as well as for a number of pulses, and to determine the
meaning of $\Phi _{\rm 0}$.  
The following notation is used:
$N(t)$ (photons cm$^{-2}$ s$^{-1}$) is the photon number flux of the
burst at time $t$, and
$N_{\rm E}(E, t)$ (photons cm$^{-2}$ s$^{-1}$ keV$^{-1}$) is 
the instantaneous spectrum of the burst at photon energy $E$ (keV)
at time $t$. 
Here,  $N_{\rm E}(E, t) dE$ is the photon number flux at time $t$
within an energy interval $dE$ around $E$.
Furthermore, 
${\cal {N}}  _{\rm E} (E) = \int _{t_1} ^{t_2} N_{\rm E}(E, t) \, dt $
(photons cm$^{-2}$ keV$^{-1}$) is the time-integrated spectrum 
of the burst and 
$E_{\rm pk}$ is the peak photon energy of $E ^2 N _{\rm E} (E, t)$
spectra.
The total photon number fluence is
$\Phi (t) =  \int _{0} ^{t}  N ( t ' ) \, dt '$ 
(cm$^{-2}$).
Note that $d \Phi (t) / dt  = N(t)$.

In \S~2, we show analytically that the exponential $E_{\rm pk}$-decay
implies that the time-integrated spectrum of a single pulse
should have a typical asymptotic
 low energy behavior of 
${\cal {N}}  _{\rm E} (E) \approx \Phi _{\rm 0}  E^{-1}$, {\it i.e.},
the same low energy power law slope as the OTTB spectrum 
(the ``universal'' $\alpha = -1$), and with the
asymptotic low energy
normalization given by  
$\Phi _{\rm 0}$. Modifications to this behavior below $E_{\rm pk,max}$
are shown to be approximately given by an incomplete gamma function. 
In \S~3, we
discuss the implications for the overall spectra of GRBs and
in \S~4, we illustrate our analytical findings by analyzing
spectral data from the bright burst, GRB 910807. 
In particular, we show that the decay constant, 
$\Phi _0$, of the spectral evolution is consistent with the 
asymptotic  normalization of the time-integrated spectrum.
Finally, in \S~5, we discuss the effects of a time-varying 
spectral shape and illustrate using GRB 950525 and GRB 921207

\section{The Time-Integrated Spectrum of a Single Pulse}

We present a scenario in which the shape of the time-integrated spectrum is 
a direct consequence, and therefore a probe of, the specific time-evolution
of the instantaneous spectra.  We assume the latter to be  
peaked in the $E N_{\rm E}$ spectrum,
{\it i.e.}, have a peak energy $E_{\rm p}$ 
around which the number of emitted photons dominate, 
which is equivalent to that
the power law indices $\alpha > -1$ and $\beta < -1$. This is indeed 
the case in most studied pulses 
(see, {\it e.g.}, Crider et~al. 1997a), except at the very end of the pulse decay, 
when slopes with 
$\alpha <-1$ sometimes are observed. This does not necessarily contradict the 
following discussion, as it can be ascribed to the possible
presence of a soft component (Preece et~al.
1996) and the contributions of previous (possibly untriggered
or unresolved) pulses.
It should also be noted that the determination of $\alpha$ 
becomes more difficult the closer the peak energy is to the low 
energy boundary of the observed spectral range. 
In any case, such a soft contribution should be negligible compared to the total 
fluence for the discussion below to be useful. 

Integrating the
instantaneous spectra over some time interval 
($t_1$ to $t_2$) of the pulse, the time-integrated spectrum becomes
\begin{eqnarray}
{\cal {N}} _{\rm E} (E) & = & \int^{t_2} _{t_1}
\NE(E, t)\,dt = \nonumber \\
& = & \int^{E_{\rm pk,min}} 
_{E_{\rm pk,max}} 
N_{\rm E}  (E, t(E_{\rm pk})) \frac{dt}{dE_{\rm pk}}\,dE_{\rm pk},
\end{eqnarray}
\noindent
where $E_{\rm pk,max} = E_{\rm pk}(t=t_1)$ and 
$E_{\rm pk,min} = E_{\rm pk}(t=t_2)$. If Equation (1) is valid during the
time interval being considered, we have
\begin{equation}
\frac{dE_{\rm pk} (t)}{dt} 
= - E_{\rm pk}(t) \frac{N(t)}{\Phi_{\rm 0}} 
= - E_{\rm pk}(t) \frac{\hat{N}(E_{\rm pk}(t))}{\Phi_{\rm 0}},
\end{equation}
\noindent
where we introduce a new function,
$\hat{N}(E_{\rm pk} (t)) =  N(t)$.
We note that the energy dependence of
$\NE(E, t ( E_{\rm pk}))$ normally is on   the
ratio $E/  E_{\rm pk}$, but that its spectral shape could be time 
dependent, adding
additional dependence on $t ( E_{\rm pk})$ or  $E_{\rm pk}$.
We therefore define a function 
$\hat{N}_{\rm E} (E/E_{\rm pk}, E_{\rm pk}) $, where
$\NE(E, t) = \hat{N}_{\rm E} (E/E_{\rm pk}, E_{\rm pk})
= \hat{N}(E_{\rm pk}) \fEEpk $, and where the function $\fEEpk$ 
describing the instantaneous spectral shape is normalized to unity, 
$\int \fEEpk\,dE = 1 $. 
Then the time-integrated spectrum, Equation (2), becomes
\begin{equation}
{\cal{N}} _{\rm E} (E) = 
 \Phi_{\rm 0} \int^{E_{\rm pk,max}} _{E_{\rm pk,min}} \fEEpk  
\frac{dE_{\rm pk}}{E_{\rm pk}}.
\end{equation}
Introducing $\fxEpk \equiv \fEEpk dE/dx = E_{\rm pk} \fEEpk$, where
$x \equiv E/E_{\rm pk}$,  and
$\fxEpk$ is normalized to unity, $\int \fxEpk\,dx = 1 $,
we find that
\begin{equation}
{\cal{N}} _{\rm E} (E) = 
\frac{ \Phi_{\rm 0}}{E} \cdot  I(E),
\end{equation}
where the correction factor
\begin{equation}
I(E) \equiv \int^{x_{\rm max} (E)} _{x_{\rm min} (E)} f _{\rm x}(x, E/x)\,dx ,
\end{equation}
and where 
$x_{\rm min} (E) = E/E_{\rm pk,max}$ and 
$x_{\rm max} (E) = E/E_{\rm pk,min}$. 
We show below that the $E$-dependence of $I(E)$ is rather weak between 
$E_{\rm pk, min}$ and $E_{\rm pk,max}$ for the cases we consider. 
The first factor in (5), describing the over-all spectral shape, shows that
the time-integrated spectrum of a single pulse,
which has an exponentially decaying peak energy, is a power law with a
time-integrated spectral index of $\ati \sim -1$ 
between $E_{\rm pk,min}$ and $E_{\rm pk,max}$
normalized by the decay constant  $\Phi_0$. This result
points out how the instantaneous and the
time-integrated spectra are related. It shows that the exponential decay
behavior of the instantaneous  spectra leaves a signature on the 
time-integrated spectrum: the decay constant, $\Phi_0$, is its asymptotic 
low energy normalization and the ``universal'' time-integrated 
$\alpha _{\rm ti} = -1$ is obtained.
The only assumption made for the derivation is the empirical
relation (1) shown so far to be valid for numerous burst pulses
(LK96, Crider et al. 1997a,c),
but whose exact validity must be explored.

Equation (1) means that a linear increase in fluence by equal steps of
$\Phi_0$ photons cm$^{-2}$ corresponds to a decrease of
$\ln E_{\rm p}$ in equal logarithmic steps.
The instantaneous spectra are, roughly, dominated by the about 
$\Phi_0$ photons cm$^{-2}$ 
in a logarithmic interval around $E = E_{\rm pk}$,
 and as a
result 
$ E {\cal N}_{\rm E} (E)$ is a constant $= \Phi_0$.
As  $ E {\cal N} _{\rm E}(E)$ also is the time-integrated
specific flux spectrum, it is clear that the photon spectrum, 
${\cal N} _{\rm E}(E)$ of a single pulse has a slope of $-1$.
The OTTB-like spectrum is thus a result
of the {\it exponential} decay of the peak energy versus photon fluence.

Let us now study the behavior of $I(E)$ when 
we can neglect the time-dependent changes of the instantaneous spectral shape,
{\it i.e}, when the $E/x$ dependence of $f _{\rm x}(x, E/x)$
in Equation (6) can be neglected. It has been shown 
for several bursts that the time-resolved spectra are well fitted 
by the Band function and therefore it is of interest to use the Band 
function in (6). However, before 
discussing these results, we study two conceptually simpler
approximations, which are valuable in illustrating the resulting behavior
of the time-integrated spectrum.
Firstly, in 
the case of integrating over all energies, {\it i.e.}, $x_{\rm min}=0$ and 
$x_{\rm max}=\infty$, $I(E)=1$ (due to the normalization of 
$f _{\rm x} (x)$) and 
$ {\cal N} _{\rm E} (E) = \Phi _{\rm 0}/E$ is an exact solution of (5). 
The shape of the 
instantaneous spectra does not affect the normalization and shape of the 
time-integrated spectrum as long as the instantaneous $E \NE(E, t)$ spectra
are peaked with the Band indices $\alpha _ {\rm i} > -1$ and
$\beta_{\rm i} < -1$. Then the dominant contribution to the integral comes from
$x\sim 1$ and 
$I(E)$ is convergent.
Secondly, if we consider finite integration limits in (6) and replace
the instantaneous spectra with 
a Dirac delta function at $E=E_{\rm pk}$, {\it i.e.}, $x=1$ and 
$f_{\rm x} (x) = \delta (x-1)$, we 
arrive at a similar result: 
$I(E) = 1$ for $E _{\rm pk , min} \leq E \leq E_{\rm pk , max}$ 
and $0$ otherwise.

\begin{figure*}
\centerline{\psfig{file=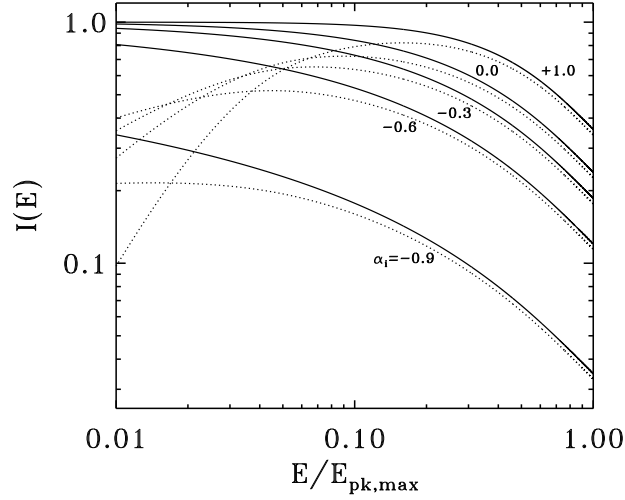,width=0.5\textwidth,angle=-0}}
\caption{The correction factor, $I(E)$, given by Eq. (6) plotted 
as a function of 
$E/E_{\rm pk,max}$ for different values of the instantaneous $\ai$. 
The curves are for $f_{\rm x} (x)$ being a Band function, 
but the 
principle behavior below $E_{\rm pk,max}$
is fairly well described  by Equation (10) alone,
{\it i.e}, for  $f_{\rm x} (x)$ being an exponentially cut-off power law.
The
solid curves describe a situation with  no low energy
cut-off, $E_{\rm pk,min} = 0$. Then Eq. (10) becomes 
$I(E) = [\Gamma (\ai +1) / A] [1 - P( \ai + 1, y _{\rm min})]$.
The dotted curves include such a cut-off, assuming $E_{\rm pk,max} / 
E_{\rm pk,min} = 20$. This corresponds to using the full
Eq. (10).\label{fig1}
}
\end{figure*}

Instantaneous spectra 
broader than a delta function will introduce 
deviations from the $E^{-1}$ power law found in the previous case.
We now consider the  Band function in (6). 
Changing variables to $y \equiv E/E_0$ and 
thus $y=x(2+\ai)$
the Band function becomes
\[ f_{\rm y} (y) = \left\{ \begin{array}{ll}
           \lbrack A \rbrack ^{-1} y^{\ai} e^{-y} & \mbox{if $y < (\ai - \bi)$};\\
           & \\
           \lbrack A' \rbrack ^{-1} y^{\bi} & \mbox{if $y \geq (\ai - \bi)$} .
           \end{array} \right. \]
For $f_{\rm y}$ to be a continuous function, we have
\begin{equation}
\frac{1}{A'}=\frac{1}{A}(\ai-\bi)^{\ai-\bi}e^{-(\ai-\bi)}
 \equiv \frac{1}{A} C(\ai-\bi),
\end{equation}
\noindent
and the normalization, $\int f_{\rm y} \,dy = 1$, gives
\begin{eqnarray}
A & = & 
 \Gamma(\ai+1) [ P(\ai+1,\ai-\bi)- \nonumber \\
  & - & \frac{ e^{-(\ai-\bi)}}
{\Gamma(\ai+1) (\bi+1)} (\ai-\bi)^{\ai+1}].
\end{eqnarray}
\noindent
where 
$P(\ai +1,y)$ is the incomplete gamma function 
 (see, {\it e.g.}, 
Press et~al. 1992). 
We have three cases:\\
(i) $y_{\rm max} > (\ai -\bi)$ and $y_{\rm min} < (\ai - \bi)$ gives
\begin{eqnarray}
I(E) & = & \frac{\Gamma(\ai+1)}{A} [ P(\ai+1,\ai-\bi)-P(\ai+1,y_{\rm min}) 
\nonumber \\
& + &  \frac{C(\ai-\bi)}{\Gamma(\ai+1)(\bi+1)}(y_{\rm max}^{\bi+1}-
(\ai-\bi)^{\bi+1} ].
\end{eqnarray}
\noindent
(ii) $y_{\rm max} < (\ai -\bi)$ gives 
\begin{equation}
I(E)=\frac{\Gamma(\ai+1)}{A} 
[ P(\ai+1, y_{\rm max})-P(\ai+1,y_{\rm min}) ].
\end{equation}  
\noindent
This is the spectrum below $(\ai-\bi)E_{\rm 0}$, and it is 
independent of $\bi$.\\
(iii) $y_{\rm min} > (\ai -\bi)$ gives 
\begin{equation}
I(E)=\frac{\Gamma(\ai+1)}{A} \cdot \frac{C(\ai-\bi)}{\Gamma(\ai+1) (\bi+1)}
(y_{\rm max}^{\bi+1}- y_{\rm min}^{\bi+1}).
\end{equation}
\noindent
At $y_{\rm max} = (\ai -\bi)$ cases (i) and (ii) are identical,
and at $y_{\rm min} = (\ai -\bi)$ cases (i) and (iii) are identical.

\begin{figure*}
\centerline{\psfig{file=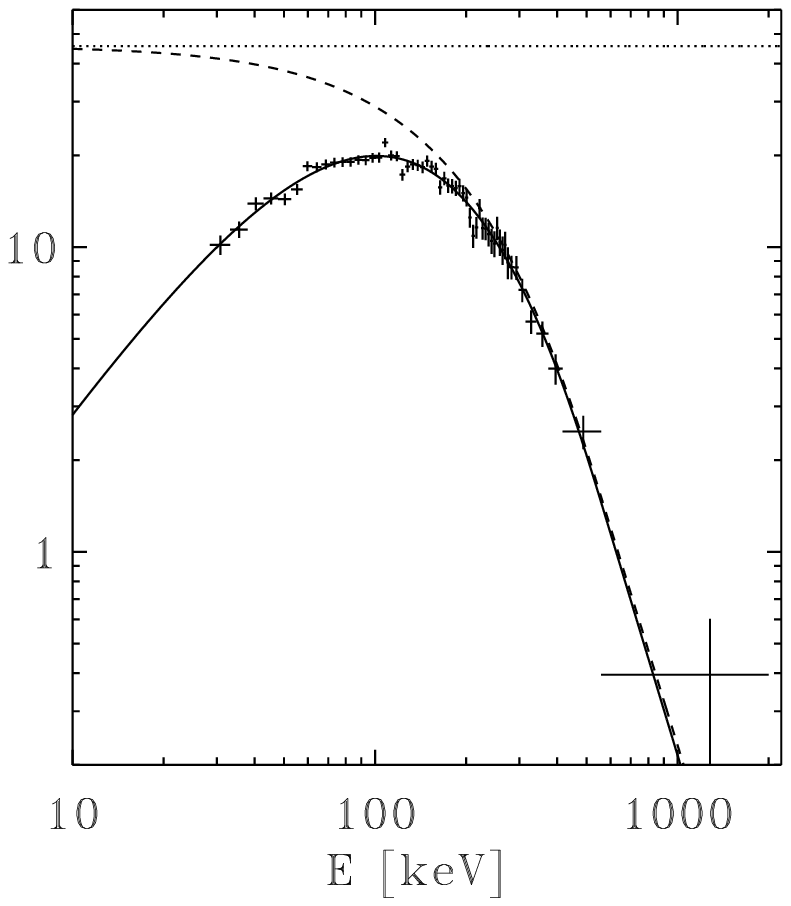,width=0.4\textwidth,angle=-0}}
\caption{Flux spectrum of the first pulse in GRB 910807 (467) and the
best fit model (solid curve); cf. Fig. 1. See the text for the fitting 
parameters.
The data were rebinned to 10 $\sigma$. The dotted line
shows the asymptotic $\ati = -1$ power law,
with the corresponding fitted normalization, $\Phi_0$.
Note that $\Phi _0$ can be directly 
read off the ordinate. The dashed curve shows the time-integrated spectrum
for the same parameters as the best fit model but with $E_{\rm pk,min} = 0$.
\label{fig2}
}

\end{figure*}

Equation (5), together with (9)-(11), relates the shape of the
time-integrated spectrum to the five parameters of the instantaneous
spectra and their time evolution, 
$\ai$, $\bi$, $\Phi_0$, $E_{\rm 0,max}$, and $E_{\rm 0,min}$. 

The general behavior of the result (9)-(11) is captured well
in an approximation where the 
instantaneous spectra are described as exponentially cut-off 
power laws, $f_{\rm y} (y) = [A]^{-1} y^{\ai} e^{-y}$, where
$A= \Gamma(\ai+1)$. 
This corresponds to a Band function without a 
high energy power law, {\it i.e.}, to the case with
$\bi = - \infty$. The result of the integration using
this  approximation has the advantage of resulting in 
the very simple analytical expression,
Equation (10), valid at all energies.
The effects on the time-integrated spectrum  due solely to a 
finite high energy cut-off energy, $E _{\rm pk,max}$, is easily seen in (10).
By letting $E _{\rm pk,min} = 0$ and thus $y_{\rm max}= \infty$,
the first term on the right-hand-side of Equation (10) equals  unity.  
Similarly, a non-zero $E_{\rm pk,min}$ will give rise to deviations from
the $E ^{-1}$-law. 
Asymptotically, at $y_{\rm max} \ll 1$, {\it i.e.}, at $ E \ll E_{\rm pk,min}$,
the time-integrated spectrum approaches 
\begin{equation}
{\cal{N}} _{\rm E} (E) = 
\frac{\Phi_0}{E_{\rm 0,min} \Gamma (\ai + {\rm 2})} \left( \frac {E} 
{E_{\rm 0,min}} \right)^{\ai},
\end{equation}
\noindent
{\it i.e.}, the time-integrated spectrum takes the low energy shape
of the instantaneous spectrum.
 
The result of integrating instantaneous spectra described by Band functions,
{\it i.e.}, Equations (9)-(11),
is shown in Figure 1. The spectra are in the form of the
correction factor, $I(E)$, and are plotted against
$E/E_{\rm pk,max}$ for different values of $\ai$.
The time-integrated 
specific flux spectra, $E \cal N \mit _{\rm E} (E)$, would be 
a constant $= \Phi _0$ in the delta function approximation, {\it i.e.},
$I(E)=1$.
Here, with a more realistic description of the instantaneous spectra,
there are deviations from the constant $\Phi _0$ in a boundary interval
below  $E _{\rm pk,max}$.
The width of this boundary
interval depends on the shape of the instantaneous spectra themselves. 
With an instantaneous photon index of $\ai = -1$, the integrated 
spectrum will never reach the asymptotic level of
$\Phi _0$. The solid curves in Figure 1 show the effect on the 
$\Phi_{\rm 0}/E$-spectrum, due to the existence of a maximal peak energy. 
The dotted curves also include the effect of a low energy bound to the peak energy.

As seen in Figure 1, the existence of a maximum and minimum peak energy makes
the time-integrated photon Band-index, $\ati$, that one would obtain
when fitting the spectrum, take a range of values enclosing the ``universal''
value of -1. This would correspond to part of the ``diversity'' of $\ati$
seen by Band et~al. (1993).
 
It is useful to introduce $E_{\rm p} (t)$ as the peak photon 
energy of the instantaneous $E N_{\rm E} (E, t)$ spectra
as the existence of such a peak
corresponds to the assumptions of the presented scenario
({\it i.e.}, $\ai > -1$ and $\bi < -1$).
In the case of the Band function, $E_{\rm pk}$ and $E_{\rm p}$ are related through
$E_{\rm pk} = E_{\rm p} (2+ \ai)/(1 + \ai)$.
The exponential decay found by LK96 holds for both $E_{\rm p}$ and $E_{\rm pk}$ as
they used a constant $\ai$ in their study.

The total time-integrated specific flux (keV cm$^{-2}$) of a pulse, 
${\cal{F}}  = \int _{0} ^{\infty} E {\cal{N}}_{\rm E} (E) \, dE $,
is dominated by the $\Phi_0$ photons cm$^{-2}$
in a logarithmic interval near
$E_{\rm p,max}$, and we thus expect 
${\cal{F}} \approx \Phi_0 E_{\rm p,max}$. Indeed, performing the
analytical integrations over $E$, using ${\cal{N}}_{\rm E} (E)$ 
for the delta function case
and the exponentially cut-off power law case, both give  
${\cal{F}} = \Phi_0 (E_{\rm p,max} - E_{\rm p,min})
\approx \Phi_0 E_{\rm p,max}$. A similar result is obtained for the Band
function case when the time-integrated $\bti \ll -2$, 
but as $\bti$ approaches -2, the
high energy power law makes an increasing and non-negligible contribution.  
The total energy of a pulse coming from an isotropic GRB at a distance $d$
is approximately
given by $4 \pi d^2 \Phi_0 E_{\rm p,max}$, making
$\Phi_0$ a useful quantity for estimating global energetics.

\section{Multi-Pulse Spectra of GRBs}

The derived time-integrated spectrum, Equation (5), 
is for a single pulse within the GRB. 
Most GRBs, however, consist of several overlapping or resolved pulses.
If the $E_{\rm pk,min}$-values of all pulses are  below the minimum energy
of the observed spectrum, then all pulses should have a similar 
low energy slope according to our  derivation, but their 
$E_{\rm pk,max}$ and
$\Phi _0$ may differ.  The sum of a number 
of time-integrated pulse spectra will lead 
to a total spectrum with two spectral ``breaks'', one at the
$E_{\rm pk,max}$ of the individual pulse with the smallest $E_{\rm pk,max}$, 
and the second at the largest $E_{\rm pk,max}$.  The 
time-integrated low energy spectrum of $k$ pulses in a burst
is the sum of the individual pulse spectra,
$ {\cal N}_{\rm E,tot} (E)= \sum _{\rm k} {\cal N}_{\rm E,k} (E) =
E^{-1} \sum _{\rm k} \Phi _{\rm 0, k} I_{\rm k} (E)$
which would give $\alpha \sim -1$
below the smallest $E_{\rm pk,max}$.
This slope may, however, not be observable 
due to the 
curvature of the individual pulse spectra (see Fig. 1) and as
the smallest $E_{\rm pk,max}$ may be close to or below
the minimum energy of the observed spectrum.
Between the smallest and the largest $E_{\rm pk,max}$,
the spectrum will be a superposition of individual pulse spectra
and 
the spectral shape  will be determined by the distributions
of $\Phi_0$ and $E_{\rm pk,max}$ among the pulses.
If, for instance, $\Phi_0$ is the same for all pulses and their 
$E_{\rm pk,max}$ is evenly distributed in logarithmic energy, 
the spectrum $E^2 {\cal{N}}_E (E)$ will be constant, 
{\it i.e.}, the photon index will be
$\sim -2$. Above the largest $E_{\rm pk,max}$, the spectrum will show the
high energy behavior of individual pulses.

The spectral diversity seen when fitting time-integrated
spectra of GRBs (Band et al. 1993) is therefore expected to mainly be 
due to the different statistical distributions of $\Phi_0$ and $E_{\rm pk,max}$
in different GRBs. The spectra of individual pulses, on the other hand,
are expected to have a spectral shape similar to the ones shown in Figure 1.
Simple GRBs with few pulses may have bumpy spectra
(possibly producing previously studied line features, see, 
{\it e.g.}, Pendleton et~al. 1994b), while complex GRBs with many pulses
will have smooth spectra.

\begin{figure*}
\centerline{\psfig{file=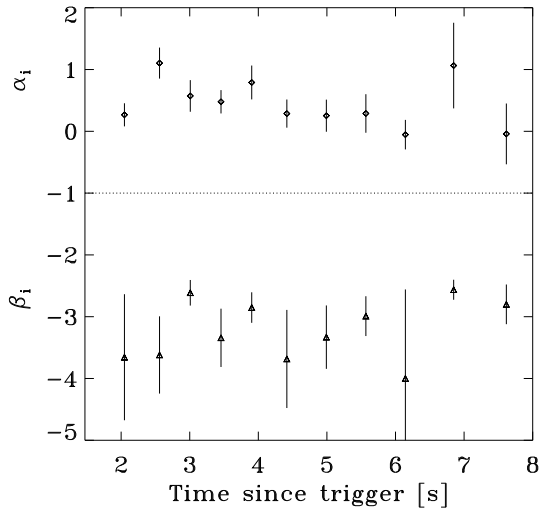,width=0.4\textwidth,angle=-0}}
\caption{The time-evolution of the shape of the instantaneous
spectra of the first pulse in GRB 910807 (467), 
described by the Band parameters 
$\alpha_{\rm i}$ and $\beta_{\rm i}$. 
The dotted line indicates the power law index $-1$.
Note that there is only a  
weak time dependence.
\label{fig3}}
\end{figure*}

\section{GRB 910807}

The spectra plotted in Figure 1 resemble real GRB pulse spectra. 
To illustrate this and the discussion above, we wish to study
a bright burst, exhibiting a smooth, long, single pulse, as 
well separated as possible from other pulses.
Generally, the 
first pulse in a burst of several separated pulses should be least 
contaminated by soft contributions  from previous pulses. 
There are only a few bursts that satisfy these  requirements, 
{\it e.g.}, the first 
pulse in the bright GRB 910807 (trigger numbers 647; 3rd BATSE catalogue, 
Meegan et~al. 1996).
Furthermore, this burst has very hard time-resolved spectra and 
no significant evolving spectral shape.
The spectra only evolve in the sense that the peak energy decreases. 
This motivates the use of our approximation in our 
deduction of the analytical expressions for the time-integrated spectrum,
in which we neglected the explicit $E_{\rm pk}$-dependence.

In Figure 2, we have plotted the time-integrated spectrum of the
Large Area Detector (LAD) data for the first 
pulse of GRB 910807. It is the specific flux spectrum, {\it i.e.},
$E {\cal{N}}_{\rm E} (E)$, and therefore the $\Phi_0$ value can directly 
be read off the ordinate. The dotted line shows the constant 
$\Phi_0 $-spectrum
which  is then modified by the $I(E)$-factor.
The solid curve is the best fit of the 
analytical result (5), with $I(E)$ found by integrating the 
Band function, {\it i.e}, equations (9)-(11) above in \S 2.
The parameters of this fit can be compared to the behavior 
of the  time-resolved spectra. For instance, their 
exponential decay, described by the decay constant, $\Phi_0$ can be 
compared to the asymptotic normalization of the time-integrated spectrum.

The best fit to the time-integrated spectrum (see Fig. 2) gives the following 
parameters:
$\Phi_{\rm 0} = 46\pm 18$ cm$^{-2}$, $E_{\rm pk,max} = 300 \pm 40$ keV, 
$E_{\rm pk,min} = 110 \pm 25$ keV, $\alpha _{\rm i} =  0.43 \pm 0.26$ and 
$\beta _{\rm i} = -4.2 \pm 0.8$ with $\chi ^2 /$d.o.f.$= 113.5/113$. 
The $\alpha _{\rm i}$- and $\beta_{\rm i}$-parameters in the fit of the 
time-integrated spectrum should be interpreted as effective values, {\it i.e.},
the values they would have had in the absence of any spectral 
shape-evolution.

\begin{figure*}
\centerline{\psfig{file=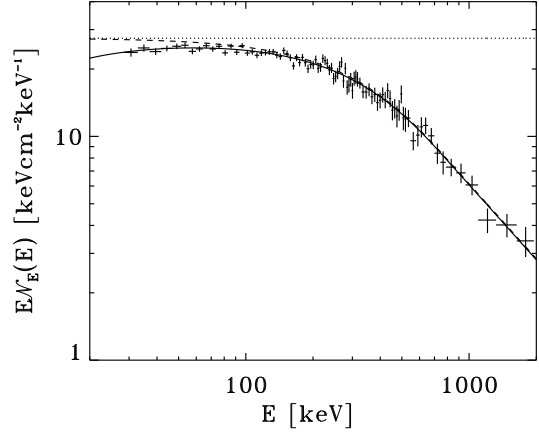,width=0.4\textwidth,angle=-0}}
\caption{Same as Fig.2., but for the first pulse in GRB 920525 (1625). 
The large energy interval over which $E_{\rm pk}$ evolves allows the 
time-integrated spectrum to reach the asymptotic level of 
$\Phi_{\rm 0}$ 
\label{fig4}
}
\end{figure*}

\begin{figure*}
\centerline{\psfig{file=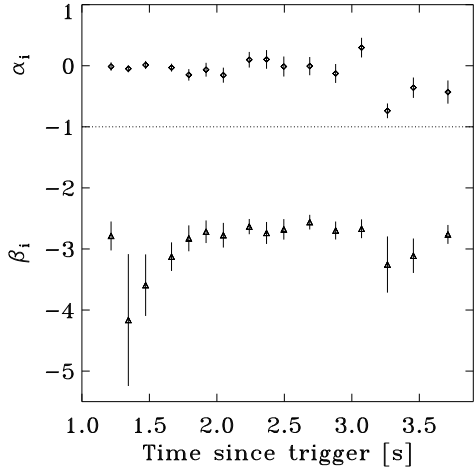,width=0.35\textwidth,angle=-0}}
\caption{Same as Fig.3., but for the first pulse in GRB 921207 (2083).
At the end of the studied time-interval, $\alpha _{\rm i}$ approaches $-1$.
\label{fig5}
}
\end{figure*}

\begin{figure*}
\centerline{\psfig{file=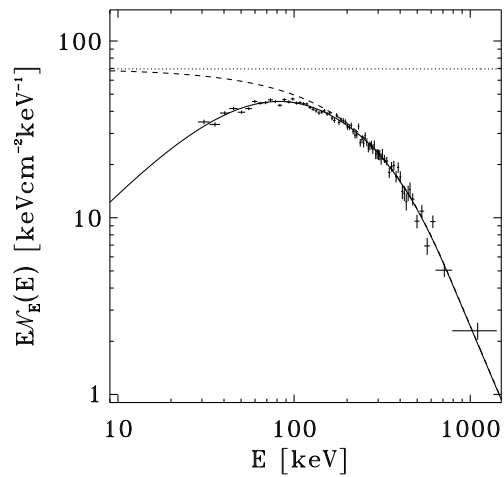,width=0.35\textwidth,angle=-0}}
\caption{Same as Fig.2., but for the first pulse in GRB 921207 (2083) 
\label{fig6}
}

\end{figure*}

The time-resolved spectra were fitted with the Band function with its 
four parameters: $\alpha_{\rm i}$, $\beta_{\rm i}$, $E_{\rm pk}$, and 
a normalization. The fits were good, having an averaged 
$\chi^2/$d.o.f.$= 104/110$. Figure 3 
presents the values of $\alpha_{\rm i}$ and $\beta_{\rm i}$ of these fits.
Note that there is no great variation in the shape of the spectra, {\it i.e.},
the $\alpha_{\rm i}$- and $\beta_{\rm i}$-parameters remain approximately
constant,
which makes the use of our approximations in \S 2 relevant.
The weighted 
averages of these two parameters are 
$< \alpha _{\rm i} >  = 0.42$ and $< \beta_{\rm i} > = -2.8$. We numerically
checked that a weak spectral shape-evolution, e.g., from $\ai = 0.6$ to 
0.2 gives a time-integrated spectrum almost identical to that 
for a constant $\ai = 0.4$. In search of $\Phi_0$,
we fit the exponential decay law (1) to the
measured $E_{\rm pk}$ vs. $\Phi$, and find $\Phi_0 = 50 \pm 4$ cm$^{-2}$.
$E_{\rm pk}$ varied from $280 \pm 20$ keV  down to $120 \pm 12$ keV. 
This is merely a factor of two, which results in that the boundary effects 
become important over the whole spectral range.

Note the expected agreement between the fit of the time-integrated
spectrum and the parameters from the fits of the time-resolved spectra.
The study above has shown that by merely studying the time-integrated 
spectrum, information about the instantaneous spectra can be extracted.
This is made possible by knowing in what way the 
time-integrated spectrum is formed from the intrinsic spectra.

\section{Effects of a Time-Varying Spectral Shape}

In the discussion above, we assumed that the explicit 
dependence of $\fxEpk$ on $E_{\rm pk}$ could be neglected, 
{\it i.e.}, the shape of the instantaneous
spectra is time-independent. This allowed us to find
illustrative analytical expressions describing the time-integrated 
spectrum. The correction to the general
time-integrated spectral shape, due to such a time dependence, is not easily 
found analytically. It should be pointed out that $\alpha _{\rm i}$ has
been observed to vary with time (see, {\it e.g.}, Crider et~al. 1997a).
It is, however, obvious that the effects on the time-integrated
spectra should be small 
for moderately varying pulse shapes, especially in the
case of the instantaneous spectra being hard.
A comprehensive study must be made numerically, in which the detailed 
time-evolution of the spectral shape is  modeled. 
Preliminary results show that the analytical expressions 
found above, {\it i.e.}, the ``averaged'' spectrum, differ 
from the numerically integrated spectra, in which $\ai$ is allowed 
to vary from its minimal to its maximal value,
by $20 - 30$ per cent or less in the most cases. 
To explore how this affects the fitted 
parameters, we have studied two bursts in which $\ai$
evolves notably and, in particular, approaches -1.
It is also of interest to study bursts, in which the peak energy 
varies over a larger range, and thus revealing more of the underlying 
$E^{-1}$- behavior, {\it i.e.}, the asymptotic normalization level.

In the first pulse of GRB 920525 (trigger numbers 1625; 3rd BATSE catalogue, 
Meegan et~al. 1996), $\alpha _{\rm i}$ varies 
between $\sim -0.3$  and $\sim -1.0$, $E_{\rm pk}$ decays from  
$1250 \pm 300$ keV to $150 \pm 70$ keV, and  
the $\Phi_{\rm 0}$-value{\footnote{LK96 found $\sim 20$ cm$^{-2}$
using LAD5, while we use LAD4, which has a stronger signal. We also use
a wider energy range. This illustrates that
the $\Phi_{\rm 0}$ results are dependent on the details of the analysis 
and therefore should be treated with some caution.}} found from the 
time-resolved evolution of $E_{\rm pk}$ is $33 \pm 1$ cm$^{-2}$.
For details of the spectral evolution, 
see, {\it e.g.}, Figure 5 in Ford et~al. (1995), 
Figure 3 in LK96, and Figure 2 in Crider
et~al. (1997a). The fact that the instantaneous
spectra are soft and evolve requires caution to be taken when the 
analytical models presented above are used.
In minimizing the $\chi ^2$ of the fit to the data, two equally 
good minima are found ($\chi ^2 /$d.o.f.$ = 97/112$), both having some
difficulties in describing the actual time-evolution. 
The first fit has $\Phi_0 = 27 \pm 2$ cm$^{-2}$ but the 
intrinsic spectra are found to be too hard; 
$\alpha_{\rm i} = +0.6 \pm 0.5$, and the $E_{\rm pk}$-values are
too small: $E_{\rm pk} = 800 - 10$ keV. 
The second fit has reasonable values for 
$\alpha _{\rm i} = -0.87 \pm 0.05$ and $\beta_{\rm i}=-2.15 \pm 0.13$, while
the rest of the parameters, although consistent with the time-resolved
fits, are poorly constrained:
$\Phi _0 = 230 \pm 220$ cm$^{-2}$, $E_{\rm pk, max} = 930 \pm 330$ keV and 
$E_{\rm pk, min} = 280 \pm 140$ keV.  
In judging these results, it should be noted that the spectra do have 
significant spectral evolution, which is not accounted for in the fitting.
Furthermore, the instantaneous spectra are quite soft, which makes
the constraining of the parameters difficult. 
Figure 4 shows the time-integrated spectrum of the first pulse in GRB 920525.
The large energy range over which the peak energy decays and 
the low value of $E_{\rm pk,min}$
allows the spectrum to reach the asymptotic level of $\Phi_{\rm 0}$.

Another example, in which $\alpha_{\rm i}$ evolves to 
soft values is the first pulse in GRB 921207 (trigger numbers 2083; 
3rd BATSE catalogue, Meegan et~al. 1996;
for details, see Figure 5 in  Ford et~al. 1995, and Figure 1 
in Crider et~al. 1997c),
with  $\alpha _{\rm i}$ varying from $\sim 0.0$ to 
$\sim -0.7$ (as shown in Fig. 5), 
$E_{\rm pk}$ decaying from $470 \pm 25$ keV to  $105 \pm 5$ keV and with 
a $\Phi_{\rm 0}$-value of $64\pm 2$ cm$^{-2}$ found from fitting
the time-resolved exponential decay of $E_{\rm pk}$. 
As shown in 
Figure 5 the $\alpha _{\rm i}$ parameter is constant in the beginning of 
the evolution but approaches $\alpha _{\rm i} = -1$ at the end, which 
could cause 
problems for the constant-$\alpha _{\rm i}$-approach. 
The time-integrated fit (see Fig. 6), however, correspond quite well to the 
time-resolved results. The fit to the time-integrated spectrum 
gives $\alpha_{\rm i} = 0.0 \pm 0.2$, $\beta _{\rm i} = -3.4 \pm 0.2$,
$E_{\rm pk,max}=505 \pm 15$ keV and 
$E_{\rm pk,min}= 68 \pm 12$ keV, and an 
asymptotic normalization of $\Phi _{\rm 0} = 70 \pm 10$ cm$^{-2}$.
The relatively good agreement between the time-integrated and the 
time-resolved fits could suggest that the spectral shape actually 
is constant in time, or at least that this is a good approximation, 
and that the late softening is caused by erroneous fits due to 
the low energy power law disappearing beyond the 
observed energy range.
Note that Crider et~al. (1997c) suggest a break in the exponential decay.

\section{Discussion}

We have presented a scenario in which the exponential decay behavior 
of the peak energy vs. fluence leads to a time-integrated
spectrum with an underlying $E^{-1}$-behavior. This spectral behavior
is not necessarily directly visible in a real spectrum, as it is 
modified by various effects, such as the existence of 
upper and lower
cut-off energies, $E_{\rm pk,max}$ and  
$E_{\rm pk,min}$. We were able to treat these effects analytically.
Other effects, such as a time-varying
spectral shape, require numerical treatment and is left for future work.

We derived a five parameter expression for the time-integrated spectrum.
These parameters describe the properties of the instantaneous spectra and 
their time evolution. Fitting a time-integrated spectrum with this 
expression thus gives information about the time-resolved spectra. We 
have illustrated our scenario by analyzing spectra from three GRB pulses. 
In future work we will consider a larger sample of pulses.
 
Studies of GRB spectra, in particular, those of individual pulses
are expected eventually to lead to an understanding
of the physical processes responsible for the GRB emission.
We emphasize that it is the instantaneous spectra and their time
evolution that reflect these processes, rather than the
time-integrated spectrum. Unfortunately, most theoretical spectral
models assume that it is the time-integrated spectrum that reflects 
the underlying physical emission mechanism.

\acknowledgments

We are grateful to Jerry Bonnell for his generous support during our
work with the BATSE data and the Wingspan program, and for his hospitality
during FR's visit to Goddard Space Flight Center. 
Furthermore, we are grateful to David Band,
Andrei Beloborodov, Claes-Ingvar Bj{\"o}rnsson, Gunnlaugur Bj{\"o}rnsson,
 Anthony Crider, Juri Poutanen and Boris Stern for 
various help.
We acknowledge support from the Swedish Natural Science Research Council
(NFR) and from a Nordic Project grant at Nordita. FR is indebted to the
Royal Swedish Academy of Sciences for travel support.

\end{document}